\documentclass[12pt]{iopart}
\usepackage{graphicx}

\begin{document}

\title{Lateral Casimir-Polder force with corrugated surfaces}

\author{Diego A. R. Dalvit$^{1}$, Paulo A. Maia Neto$^{2}$, Astrid Lambrecht$^{3}$ and Serge Reynaud$^{3}$}

\address{$^{1}$ Theoretical Division, Los Alamos National Laboratory, Los Alamos, NM 87545, USA
\\
$^{2}$ Instituto de F\'{\i}sica, UFRJ, 
CP 68528,   Rio de Janeiro,  RJ, 21941-972, Brazil
\\
$^{3}$ Laboratoire Kastler Brossel,
CNRS, ENS, Universit\'e Pierre et Marie Curie case 74,
Campus Jussieu, F-75252 Paris Cedex 05, France
}

\begin{abstract}
We  derive the lateral Casimir-Polder force on a ground state atom 
on top of a corrugated surface, up to first order in the corrugation amplitude.
Our calculation is based on the scattering approach, which takes into account nonspecular 
reflections and polarization mixing for electromagnetic quantum fluctuations
impinging on real materials. We compare our first order exact result with two commonly
used approximation methods. We show that the proximity force approximation (large corrugation
wavelengths) overestimates the lateral force, while the pairwise summation approach 
underestimates it due to the non-additivity of dispersion forces. 
We argue that a frequency shift  measurement for the dipolar lateral oscillations of cold atoms could provide a striking demonstration of nontrivial  geometrical effects on the quantum vacuum. 
  
\end{abstract}

\submitto{\JPA}

\maketitle

\section{Introduction}

The quantum vacuum electromagnetic field plays an essential role in the interaction between ground state atoms \cite{CasimirPolder1948}. 
These van der Waals/Casimir-Polder interactions also take place for an atom near a 
bulk surface or between two bulk surfaces \cite{Casimir1948} (Casimir effect) \cite{Milonnibook}.  
Several methods have been employed to measure the atom-surface Casimir-Polder interaction 
\cite{Hinds1993,Aspect1996,Shimizu2001,DeKieviet2003,Vuletic2004,Ketterle2004,Ketterle2006,
Cornell2005,Cornell2007}.
The Casimir force between bulk surfaces has been measured for
different geometries  over the last years \cite{Lamoreaux1997,Mohideen1998,Ederth2000,Bressi2002,Decca2003}, 
paving the way for promising applications in nanotechnology \cite{Capasso2001}.

On the theoretical front, surface geometries are usually analyzed with the help of the 
proximity force approximation (PFA, also known as Derjaguin approximation) \cite{Derjaguin}.   
In this approach, the interaction energy between curved surfaces is obtained from the plane-plane case by a simple average over local distances (see, for instance, \cite{Israelachvili}). 
PFA-based theories have been employed to compare with experimental results for the 
normal Casimir force between a plane and a sphere \cite{Lamoreaux1997,Mohideen1998, Decca2003}, between crossed cylinders \cite{Ederth2000}, between a plane and a cylinder 
\cite{Onofrio2005}, as well as for the lateral Casimir force between corrugated plates \cite{Mohideen2002}. 

However, PFA  only holds when the separation distance is much smaller than the length scales characterizing the surface geometries (radius of curvature for spherical and cylindrical plates, and corrugation wavelength for corrugated surfaces). Thus, 
it is very important to check the accuracy of PFA predictions by comparing them with more rigorous formalisms. Beyond-PFA theories for normal Casimir force for cylindrical 
\cite{nonPFAcylindersEmig2006,nonPFAcylindersDalvit2006} and 
spherical \cite{nonPFAspheres} plates, and for the lateral Casimir force 
\cite{Buscher2005,nonPFAlateral} have been recently reported. Most of these theories, though, cannot be directly compared to existing experimental data, either because of the assumption of perfect reflectivity \cite{nonPFAcylindersEmig2006,nonPFAcylindersDalvit2006,Buscher2005} (a very poor
approximation for the typical separation distances of a few hundred nanometers employed in the experiments), or because  assumptions about the geometry itself (small corrugation amplitudes)  \cite{nonPFAlateral} are not satisfied by the actual experimental conditions \cite{Mohideen2002}.

In this paper, we propose that the lateral Casimir-Polder force on a ground-state atom on top of a 
corrugated surface provides a striking evidence of a non-trivial ({\it i.e.} beyond PFA) geometry effect. In fact, we show that a Bose-Einstein condensate (BEC) trapped close to a corrugated surface may be used as a local field sensor of the geometry-induced reshaping of quantum vacuum fluctuations.  The control of the atom-surface interaction may be relevant for
several applications involving cold atoms near surfaces \cite{reviewatomsurface}.
Deviations from the PFA regime are considerably larger than in the case of the lateral force for bulk surfaces, where the geometry effect is averaged over the surfaces. 

We employ the scattering approach \cite{NJP2006} and write the interaction energy in terms of 
a non-specular reflection operator that takes into account diffraction and polarization mixing. 
A similar formalism was developed previously to analyze the roughness correction to the normal Casimir force \cite{EPL2005} and to the lateral Casimir force between corrugated bulk plates 
\cite{nonPFAlateral}. The corrugation is taken into account as a perturbation of the plane geometry, and the interaction potential is computed up to first order in the corrugation 
amplitude $a$. Hence our results only hold when $a$ is the smallest length scale in the problem: $a\ll z_A,\lambda_c$, but arbitrary relative values for the atom-surface distance $z_A$ (measured with respect to some arbitrary reference plane) and the corrugation wavelength $\lambda_c$ are allowed. Our results are valid for any real material, not necessarily a perfect conductor, and exact to first order in perturbation theory in powers of the corrugation amplitude. 

We recover the PFA results when taking $z_A \ll \lambda_c$. In this limit, the effect is dominated by specular reflections, since the the surface is nearly plane in the scale of the separation distance. The resulting PFA potential is trivially connected to the plane case by taking the local atom-surface distance. 
We also compare our results with those obtained from the pairwise summation (PWS) approach developed in \cite{Klimchitskaya2000}. Since the Casimir-Polder force is actually not additive 
\cite{Milonnibook}, Ref. \cite{Klimchitskaya2000} corrects  the result of the pairwise integration by a pre-factor ``calibrated" by comparing the pairwise and exact results for the 
perfectly reflecting, planar geometry in the retarded Casimir-Polder regime ($z_A \gg \lambda_A$, 
where $\lambda_A$ is the typical wavelength characterizing the optical response of the atom).
As could have been expected, the resulting lateral force coincides with the more rigorous results presented in this paper when considering nearly plane corrugated surfaces, $z_A \ll \lambda_c$.
In this  limit, both our exact result and the approximate PWS approach agree with the 
approximate PFA approach. Outside the PFA regime, however, PFA overestimates and PWS underestimates the lateral effect, the discrepancy increasing as the corrugation wavelength gets shorter. 

In the present paper, we focus on the comparison between our results and the PFA and PWS approaches. For simplicity, we take a perfectly-reflecting surface for the  numerical examples, 
leaving the more general discussion to a future publication \cite{future}. A short summary of our results, including the discussion of real materials, was reported by us elsewhere \cite{arxiv}.  


\section{Casimir-Polder potential for a corrugated surface}

The corrugated surface is described by the profile function  $h({\bf r})$ (with ${\bf r} = (x,y)$) giving the local height with respect to the reference plane $z=0$. The potential energy of a ground-state atom located at some position ${\bf R}_A=(x_A, y_A, z_A)$ near the surface
is written as
\begin{equation}
U({\bf R}_A) = U^{(0)}(z_A) + U^{(1)}({\bf R}_A).
\end{equation}
The zeroth-order potential $U^{(0)}(z_A)$, which contains only specular reflection amplitudes, is the potential energy for an atom above a plane plate at $z=0$.
We make a perturbative expansion in powers of $h$ and use the scattering approach to find the first-order correction $U^{(1)}$.  The first-order scattering by the corrugated surface is described by the matrix elements of the first-order reflection operator:
$\langle  {\bf k}, p | {\cal R}^{(1)} | {\bf k}', p' \rangle = R_{p,p'}({\bf k},{\bf k}') H({\bf k}-{\bf k}')$,
where $H({\bf k})$ is the Fourier transform  of the profile function $h({\bf r})$.
We find
\begin{equation} 
U^{(1)}({\bf R}_A) = \int \frac{d^2 {\bf k}}{(2\pi)^2} \, e^{i {\bf k} \cdot {\bf r}_A}  \, 
g({\bf k}, z_A) \, H({\bf k}).
\label{lateralenergy}
\end{equation}
To first order in the corrugation amplitude this expression is exact, and
depends on the response function $g({\bf k}, z_A)$ which can be explicitly written in terms of the 
first-order  nonspecular reflection coefficients \cite{EPL2005}
$R_{pp'}({\bf k}, {\bf k}')$ as
\begin{equation} \label{g}
g({\bf k}, z_A) =  \frac{\hbar}{c^2 \epsilon_0} \int_0^{\infty} \frac{d\xi}{2\pi} \;  \xi^2 \alpha(i \xi)
\int \frac{d^2{\bf k}'}{(2\pi)^2} a_{{\bf k}', {\bf k}'-{\bf k}}(z_A,\xi) , 
\end{equation}
\begin{equation} \label{a}
a_{{\bf k}, {\bf k}'}(z_A,\xi) = \frac{1}{2\kappa'}\sum_{p,p'} 
\mbox{\boldmath\({\hat \epsilon}\)}_{p}^{+}\cdot
\mbox{\boldmath\({\hat \epsilon}\)}_{p'}^{-}  e^{-(\kappa+\kappa')z_A} \,
R_{pp'}({\bf k}, {\bf k}') ,
\end{equation}
where $\alpha(i \xi)$ is the dynamic atomic polarizability of the atom
(whose ground state is assumed to be spherically symmetric), and $\epsilon_0$ is the
vacuum permittivity. Here
${\bf k}$ and ${\bf k}'$ are the lateral wave vectors of the input and output
fields, and $p$ and $p'$ are their polarizations (TE for transverse electric and TM
for transverse magnetic). The unit polarization vectors for outgoing fields
are \(\mbox{\boldmath\({\hat \epsilon}\)}^{+}_{\rm TE}({\bf k}) = {\bf z} \times {\bf k}\)
and \(\mbox{\boldmath\({\hat \epsilon}\)}^{+}_{\rm TM}({\bf k}) = 
\mbox{\boldmath\({\hat \epsilon}\)}^{+}_{\rm TE} \times {\bf K}\), with
${\bf K} = {\bf k} + k_z {\bf z}$, and $k_z = {\rm sgn}(\omega) \sqrt{\omega^2/c^2 - k^2}$.
Similar expressions hold for the polarization vectors 
\(\mbox{\boldmath\({\hat \epsilon}\)}^{-}_p({\bf k})\) for incoming fields, with the replacement
$k_z \rightarrow - k_z$.  The roundtrip propagation of the field between the surface and the atom
is contained in the factor $\exp[-(\kappa+\kappa') z_A]$, with 
$\kappa=\sqrt{\xi^2/c^2 + k^2}$. 

Using that the atomic polarizability along the imaginary frequency
axis is real and positive, one can prove that the  response function $g({\bf k}, z_A)$  is explicitly real.
Moreover, as expected from symmetry, the  response function does not depend on the direction of 
the corrugation profile: $g({\bf k}, z_A) = g(k,z_A).$ 


\section{Proximity force approximation}

The response for very smooth surfaces is given by the limit $k\rightarrow 0$ in Eq.~(\ref{g}). According to (\ref{a}),  $g(0,z_A)$ is given entirely in terms of the specular limit of the first-order nonspecular coefficients, which quite generally satisfy \cite{EPL2005}
\begin{equation}\label{specular}
R_{pp'}({\bf k},{\bf k})= 2\kappa r_p({\bf k}) \delta_{p,p'},
\end{equation}
with $r_p({\bf k})=r_p(k)$ representing the standard Fresnel coefficients for a plane surface. 
This result has a simple  interpretation: in the specular limit, the first-order correction is just the additional roundtrip phase $2 i k_z h$ associated to the surface displacement $h.$ After replacing 
(\ref{specular}) into (\ref{a}), we are able to connect the resulting expression for $g(0,z_A)$ with the (zeroth-order) potential for a plane surface:
\begin{equation} \label{PFA}
g(k_c\rightarrow 0,z_A) = - \frac{{\rm d} U^{(0)}(z_A)}{{\rm d}z_A} .
\end{equation}
It follows from the ``proximity force theorem" (\ref{PFA}) that the potential for a very smooth surface 
can be obtained from the planar case by merely taking the local atom-surface distance:
\begin{equation}
U({\bf R}_A) \approx U^{(0)}(z_A-h({\bf r}_A)) \approx U^{(0)}(z_A) - h({\bf r}_A) 
\frac{{\rm d} U^{(0)}(z_A)}{{\rm d} z_A} .
\end{equation}
This ``proximity force approximation" holds when $g(k,z_A)$ may be replaced by $g(0,z_A)$ for all 
Fourier components of the surface profile function $H({\bf k})$ contributing appreciably in Eq.~(\ref{lateralenergy}).
Deviations of the PFA from the exact result (for a given Fourier component) are measured by the ratio 
\begin{equation}\label{rho}
\rho_{\rm PFA} = \frac{g(k,z_A)}{g(0,z_A)}.
\end{equation}


\section{Pairwise summation approximation}

The pairwise summation (PWS) technique is an approximation different from PFA that computes the
Casimir force between bulk surfaces or the Casimir-Polder force between an atom and a bulk surface
by addition of the atom-atom interaction potential, which is 
$U(r_{12})= -(23/ 4 \pi) \alpha_{A_1}(0) \alpha_{A_2}(0) \; r_{12}^{-7}$ for two ground-state atoms separated by a distance $r_{12}$ much larger than the static atomic polarizabilities (retarded regime). 
The PWS method has been discussed in \cite{Emig2003} for the normal and lateral Casimir forces between corrugated, perfectly reflecting bulk plates, and has been used to 
calculate lateral CP \cite{Klimchitskaya2000} and vdW forces \cite{HenkelPhD} 
between an atom and a rough or corrugated bulk plate. 

The pairwise lateral Casimir-Polder energy for an atom above a perfectly reflecting surface is given by
\begin{equation}
U_{\rm PWS}({\bf R}_A) = - {\cal C} \int d{\bf r} \int_0^{h({\bf r})} dz 
\frac{1}{[({\bf r}-{\bf r}_A)^2 + (z-z_A)^2]^7} ,
\end{equation}
where ${\cal C}= 15 \hbar c \alpha(0) /8 \pi^2 \epsilon_0$ is a ``calibration" factor (sometimes improperly called ``normalization" factor) chosen such that the PWS result coincides with the exact known
result for a planar surface. To first order in the corrugation amplitude, the lateral PWS potential
can be written in a form similar to our exact result (\ref{lateralenergy})
\begin{equation}
U^{(1)}_{\rm PWS}({\bf R}_A) =  \int \frac{d^2{\bf k}}{(2 \pi)^2} e^{i {\bf k} \cdot {\bf r}_A} g_{\rm PWS}({\bf k}, z_A) H({\bf k}) ,
\label{lateralenergyPWS}
\end{equation}
where the PWS response function is
\begin{eqnarray}
g_{\rm PWS}({\bf k}, z_A) &=& - {\cal C} \int d{\bf r} \frac{e^{-i {\bf k} \cdot {\bf r}}}{[{\bf r}^2 + z_A^2]^{7/2}}  =
- \frac{3 \hbar c \alpha(0)}{8 \pi^2 \epsilon_0 z_A^5} F_{\rm PWS}(k z_A) , \nonumber \\
F_{\rm PWS}({\cal Z}) &=& e^{- {\cal Z}} (1 + {\cal Z} + {\cal Z}^2/3) .
\end{eqnarray}
Again, due to symmetry, the PWS response function depends only on the modulus of ${\bf k}$. 
Deviations of the PWS from the exact result (for a given Fourier component) can be measured by the ratio
\begin{equation}
\rho_{\rm PWS} = \frac{g(k, z_A)}{g_{\rm PWS}(k, z_A)} .
\end{equation}
Clearly, in the limit of long corrugation wavelengths $k z_A \ll 1$, both PFA and PWS coincide with the exact
result. However, as the corrugation wavelength decreases, both approximations depart from the exact result
given in Eq.(\ref{lateralenergy}).

The use of a ``calibration" factor ${\cal C}$ is a major uncontrolled approximation of the PWS technique, as it is chosen such that the PWS energy coincides with the exact one for a particular distance (in the above, for the retarded Casimir-Polder regime) and for a particular model for the surfaces (above, perfect conductors).
Clearly, such a uniquely fixed calibration factor cannot account for the different geometry and materials dependency
of dispersion forces in various distances regimes. In view of the non-additivity of Casimir forces, the pairwise summation technique is therefore an uncontrolled approximation only valid for very dilute media for restricted
length scales.


\section{Sinusoidal corrugation}

In this Section we will concentrate on sinusoidal corrugated surfaces, and later on we will
consider a grooved surface where non-trivial (beyond-PFA) effects of geometry are even more impressive.

For a uni-axial sinusoidal corrugation $h(x) = h_0 \cos(k_c x)$
we find from (\ref{lateralenergy})
\begin{equation} \label{sinusoidal}
U^{(1)}(x_A, z_A) = h_0\,g(k_c,z_A)\,\cos(k_c x_A).
\end{equation}
Since $g$ is negative (see below), this potential brings the atom to one of the surface crests. 
For simplicity, we only consider the case of perfect reflectors (see Refs.~\cite{arxiv} and \cite{future} for
a more general discussion of real materials). In this case, the first-order reflection coefficients have a very simple 
form \cite{EPL2005}
\begin{eqnarray}
R_{{\rm TE}, {\rm TE}}({\bf k}, {\bf k}') = -2 \kappa' C \; &;& \; 
R_{{\rm TE}, {\rm TM}}({\bf k}, {\bf k}') = -2 \xi S / c ; \nonumber \\
R_{{\rm TM}, {\rm TE}}({\bf k}, {\bf k}') = -2 \frac{\xi \kappa'}{c \kappa} S \; &;& \; 
R_{{\rm TM}, {\rm TM}}({\bf k}, {\bf k}') =  \frac{2 k k'}{\kappa} + \frac{2 \xi^2}{c^2 \kappa} C ,
\end{eqnarray}  
with $C=\cos(\phi-\phi')$ and $S=\sin(\phi-\phi')$ representing respectively the cosine and the sine of the
angle between ${\bf k}$ and ${\bf k}'$.
These expressions allow us to derive some simple analytical results for the exact response function $g(k, z_A)$.
In the non-retarded van der Waals regime ($z_A\ll \lambda_A$)
\begin{eqnarray}
g(k_c, z_A) &=& - \frac{\hbar G(k_c z_A)}{64 \pi^2 \epsilon_0 z_A^4} \int_0^{\infty} d\xi \; \alpha(i \xi) ,
\nonumber\\
G({\cal Z}) &=& {\cal Z}^2 [ 2 K_2({\cal Z}) + {\cal Z} K_3({\cal Z})], 
\label{vdWperfect}
\end{eqnarray}
where, as usual, the vdW energy depends on the whole frequency spectrum of dynamic atomic polarizability 
along the imaginary frequency axis. $K_2$ and $K_3$ are the modified Bessel functions of second and third order, respectively \cite{Abramowicz}.
In the retarded Casimir-Polder regime  ($z_A \gg \lambda_A$)
\begin{eqnarray}
g(k_c, z_A) &=& - \frac{3 \hbar c \alpha(0)}{8 \pi^2 \epsilon_0  z_A^5} F(k_c z_A), 
\nonumber\\
F({\cal Z}) &=& e^{-{\cal Z}} (1+{\cal Z}+ 16 {\cal Z}^2/45 + {\cal Z}^3/45).
\label{CPperfect}
\end{eqnarray} 
As usual, the CP expression depends only on the static atomic polarizability.
Both analytical results are consistent with the proximity force theorem (\ref{PFA}) and
deviations from PFA are given by $\rho_{\rm PFA} = G(k_c z_A)/12$ for the vdW regime, 
and $\rho_{\rm PFA} = F(k_c z_A)$ for the CP regime. Eq.~(\ref{CPperfect}) has a greater practical interest because the assumption of perfect reflectivity is better suited at longer separation distances. In this Casimir-Polder regime, deviations from PWS are given by
$\rho_{\rm PWS} = F(k_c z_A) / F_{\rm PWS}(k_c z_A)$.

\begin{figure}[b]
\centering
\includegraphics[width=6cm]{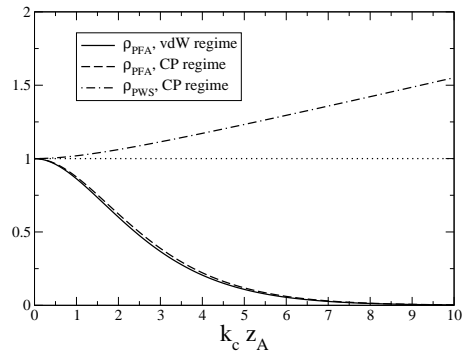}
\caption{Comparison between the first order, exact result with the proximity force approximation
and the pairwise summation approach for the lateral Casimir-Polder interaction between an atom and a
perfectly conducting, sinusoidal corrugated surface.}
\end{figure}

In Fig. 1 we plot the deviations of the proximity force approximations and of the pairwise
summation approximation from the exact result. For PFA we plot the analytical curves for
the vdW and CP regime, while for PWS we only plot the curve for the CP regime. Since the
calibration constant ${\cal C}$ for the PWS has been calibrated for the Casimir-Polder regime, the 
pairwise summation prediction for the van der Waals regime makes no sense, and for this reason
we do not plot $\rho_{\rm PWS}$ in the vdW regime. From the figure we conclude that, for the
sinusoidal corrugation, PFA always
overestimates the lateral force, while PWS always underestimates it. To give a number on the 
deviation from the PFA and PWS, let us consider a separation $z_A=2 \mu{\rm m}$, well within
the CP regime, and a corrugation wavelength $\lambda_c = 3.5 \mu{\rm m}$ ($k_c z_A \approx 3.55$).
In this case, $\rho_{\rm PFA} \approx 30 \%$ and $\rho_{\rm PWS} \approx 115 \%$, which means
that the PFA largely overestimates the magnitude of the lateral effect, while the PWS underestimates
it by as much as $15 \%$ for this not too large value of $k_c z_A$.


\section{Grooved corrugation}

Larger deviations from PFA can be obtained for a surface with periodical grooves (Fig. 2). 
In this case the potential (\ref{sinusoidal}) has to be replaced by a sum over Fourier components $a_n$ of
the corrugation profile (assumed to be even for simplicity) $h(x)=\sum_n a_n \cos(n k_c x)$,
\begin{equation}
U^{(1)}(x_A, z_A) = \sum_{n=0}^{\infty} a_n \cos(n k_c x_A) \;  g(n k_c, z_A) ,
\label{energygrooved}
\end{equation}
where the Fourier coefficients $a_n$ are 
\begin{equation}
a_n =
\left\{ \begin{array}{ll}
a \left( 1- \frac{s}{2 \lambda_c} \right) & \mbox{for $n=0$} \\
(-1)^{n+1}\;  \frac{2 a \lambda_c}{\pi^2 s}  \; \frac{1-\cos(n \pi s / \lambda_c)}{n^2} & \mbox{for $n \ge 1$}
\end{array}\right.
\end{equation}

\begin{figure}[t]
\centering
\includegraphics[width=8cm]{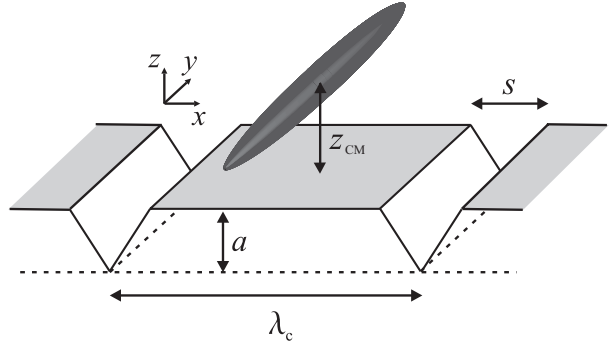}
\caption{Periodically grooved plate for measuring large corrections to the
proximity force and pairwise summation approximations. A single cold atom or an elongated
Bose-Einstein condensate (as in the figure) is trapped with an external asymmetric harmonic potential
above the plateau region. The dipole oscillation frequency along the $x$ direction is shifted due
to the lateral Casimir-Polder force.}
\end{figure}

In Fig. 3 we compare the exact, PFA and PWS lateral Casimir-Polder potentials for various values of
$k_c z_A$. The PFA potential has the same profile as the corrugation itself for all values of $k_c z_A$,
while the exact and PWS potentials coincide with the PFA version only for $k_c z_A \ll 1$.
The PFA potential is recovered when the response function $g(n k_c, z_A)$ may be replaced by the 
limiting expression $g(0, z_A)$ for all values of $n$ significantly contributing to the profile.
Otherwise, as $z_A$ increases, multiplication by $g(n k_c, z_A)$ renders the contribution
of higher orders comparatively smaller. When $k_c z_A \gg 1$, the exponential decrease of $g$ implies
that the first order $n=1$ term dominates the sum (\ref{energygrooved}) 
(apart from the irrelevant $n=0$ term that is $x$ independent and does not contribute to the lateral force), and both the exact and PWS potentials are approximately sinusoidal. In this limit,
the amplitude of the effective sinusoidal exact potential is 
\begin{equation}
h_0 = \frac{2 a \lambda_c}{\pi^2 s} \; (1-\cos(k_c s / 2)) \; g(k_c, z_A) \; ; \; [ k_c z_A \gg 1 ] 
\end{equation}
and for the PWS potential a smiliar equation holds, with $g$ replaced by $g_{\rm PWS}$. 
For given values of $z_A$, $a$, and $\lambda_c$, the amplitude is maximized for $s \approx 0.74 \lambda_c$,
which corresponds to a rather narrow plateau. It seems more realistic to take, say, $s=\lambda_c/2$,
as we do in all our numerical calculations in this paper.

When the atom is located above one plateau, the PFA predicts that the lateral Casimir-Polder force
should vanish, since the energy is thus unchanged in a small lateral displacement. A non vanishing
force appearing when the atom is moved above the plateau thus clearly signals a deviation from the
PFA. On the other hand, deviations from the PWS are not as impressive as for the PFA, but should
nevertheless be observable for not too large values of $k_c z_A$. For example, for $k_c z_A=3.55$,
PWS underestimates the amplitude of the exact potential by $15\%$, similarly as for the sinusoidal
corrugation case.

\begin{figure}[t]
\centering
\includegraphics[width=6.5cm]{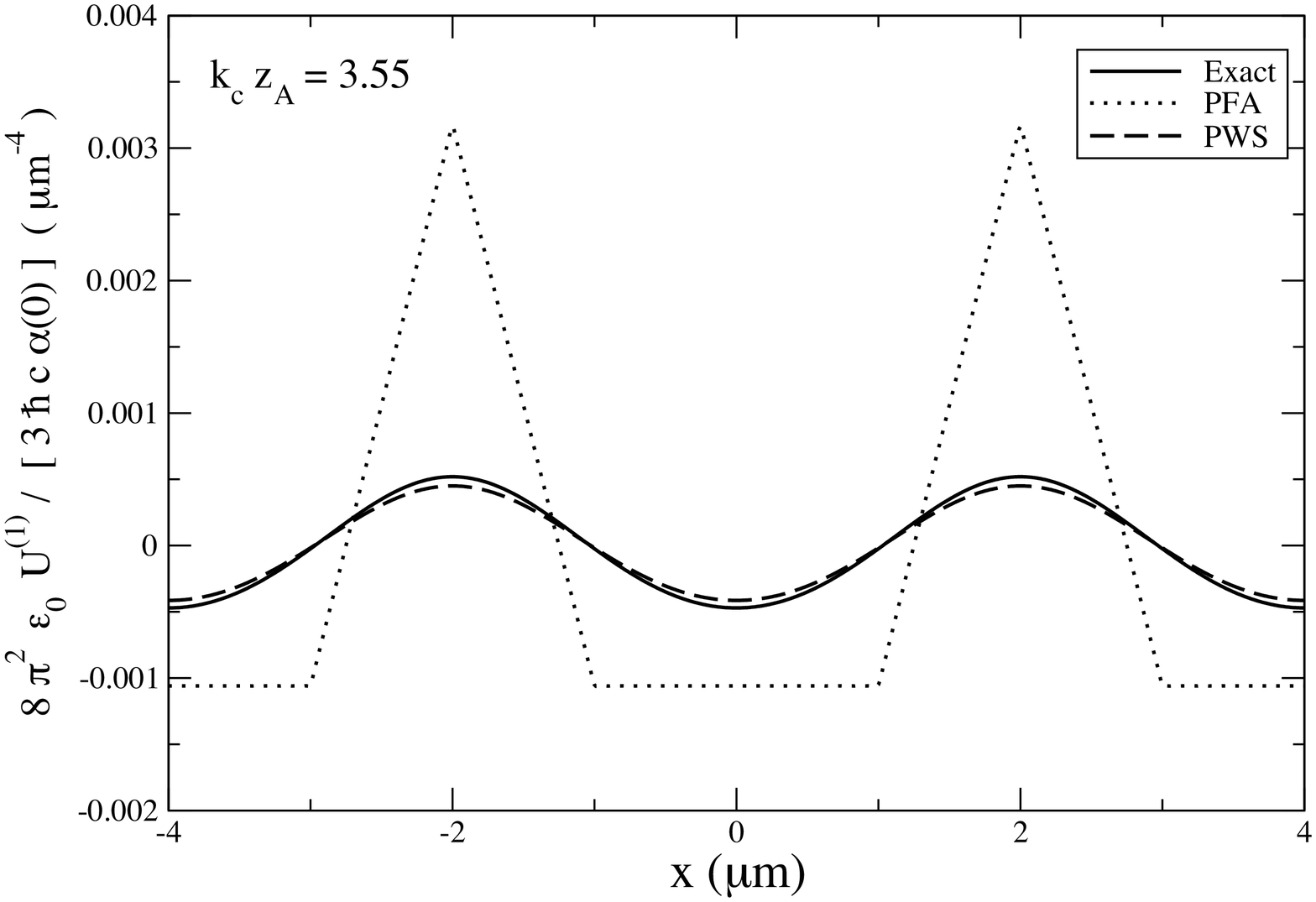}
\includegraphics[width=6.5cm]{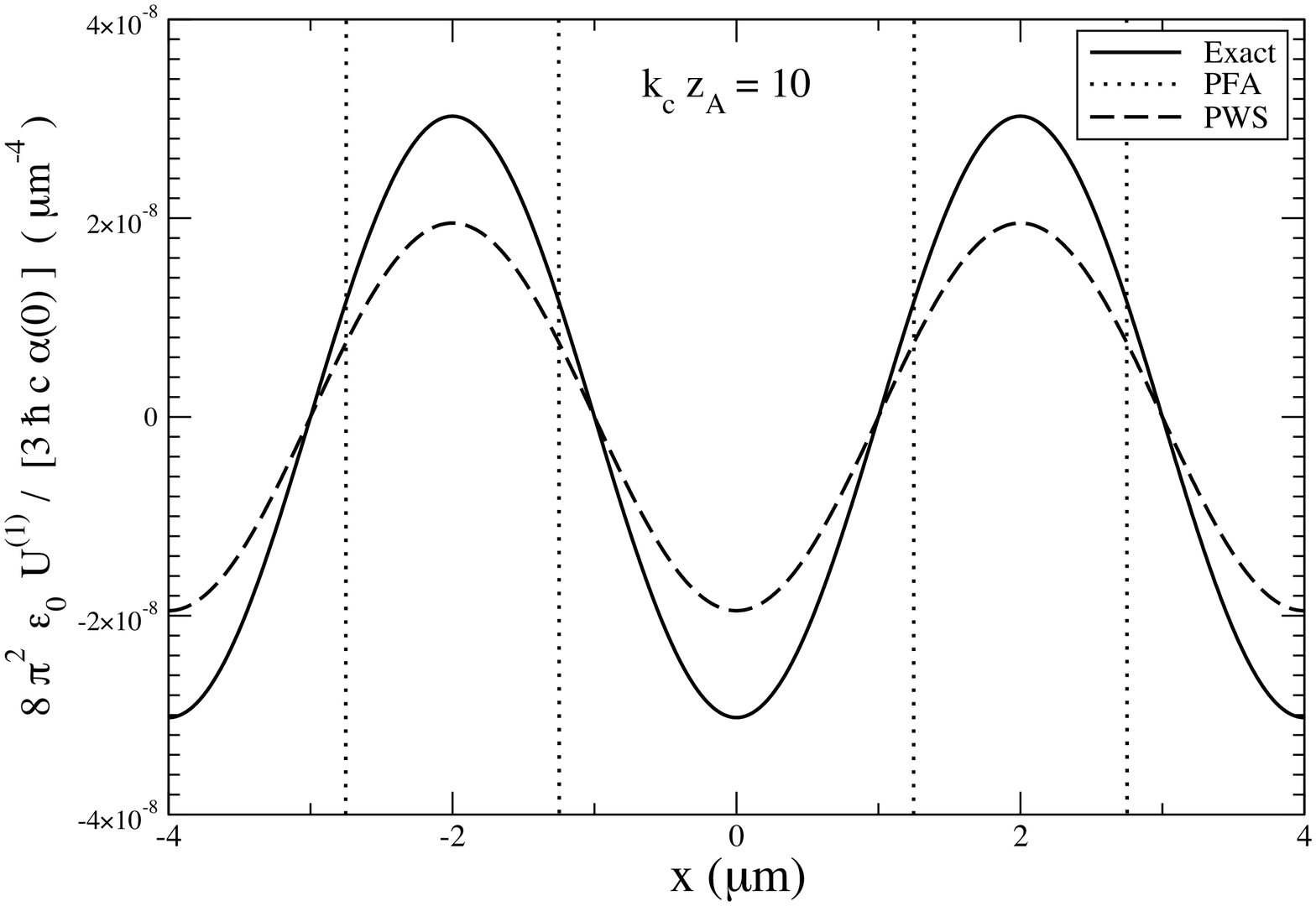}
\caption{Comparison between the exact, proximity approximation, and pairwise approximation for the
lateral Casimir-Polder potential between an atom and a perfectly conducting, periodically grooved 
surface for different values of $k_c z_A$. The groove period is $\lambda_c=4 \mu{\rm m}$, the groove width $s=\lambda_c/2$,
and the depth $a=250 {\rm nm}$. }
\end{figure}


\section{BEC as a sensor of lateral Casimir-Polder forces}

Dipolar oscillations of the center-of-mass of a Bose-Einstein condensate trapped close to a 
surface have been proposed as a highly sensitive probe of quantum vacuum fluctuations \cite{Antezza2004}. 
This idea has been successfully implemented to measure the normal component of the Casimir-Polder force
between rubidium atoms above a dielectric (fused silica) surface in the $5-10 \mu{\rm m}$ separation
range \cite{Cornell2005}, and to measure non-equilibrium temperature corrections to the Casimir-Polder
force \cite{Cornell2007}. The reported relative frequency shift sensitivity in these experiments was
$\gamma = 10^{-4} - 10^{-5}$.

For the lateral Casimir-Polder force we have in mind a setup where
the long axis of an elongated BEC would be parallel to the corrugations, while the lateral (along the
$x$ direction) dipolar oscillation of the center-of-mass would be monitored as a function of time (see Fig. 2). When a BEC trapped in an external asymmetric harmonic potential of frequencies 
$\omega_x = \omega_z \gg \omega_y$ is set to oscillate perpendicular to the grooves, the
lateral Casimir-Polder force renormalizes the center-of-mass (CM) dipolar oscillation frequency in the $x$ direction,
\begin{equation}
\omega_{{\rm CM},x}^2 = \omega_x^2 + \frac{1}{m} \int d{\bf r} \;n_0({\bf r}) \; 
\frac{\partial^2 U^{(1)}(x,z)}{\partial x^2} ,
\end{equation}
where $m$ is the atomic mass, and $n_0({\bf r}) = \frac{15}{6 \pi R^5} ( R^2 - (x^2+z^2))^{3/2}$ is the two-dimensional Thomas-Fermi density for a BEC of Thomas-Fermi radius $R$. The relative frequency shift
$\gamma = (\omega_{{\rm CM},x} - \omega_x)/\omega_x$ is given by
\begin{eqnarray}
\gamma &=& - \frac{15 k_c^2}{12 m \omega_x^2} \; \sum_{n=1}^{\infty} n^2 a_n  \\
&& \times \int_0^1 d\rho \; \rho \; (1-\rho^2)^{3/2} \int_0^{2 \pi} d\theta \; g(n k_c, z_{\rm CM} + \rho R \sin \theta) \; \cos(n k_c \rho R \cos \theta) , \nonumber
\end{eqnarray}
where $z_{\rm CM}$ is the distance from the CM of the BEC to the plate. 
The relative frequency shift $\gamma_0$ for a single atom oscillating above the corrugated surface is obtained in the limit $R \ll z_{\rm CM}, \lambda_c$, namely
\begin{equation}
\gamma_0 = - \frac{k_c^2}{2 \omega_x^2 m} \; \sum_{n=1}^{\infty} n^2 \; a_n \; g(n k_c, z_A) .
\end{equation}

\begin{figure}[b]
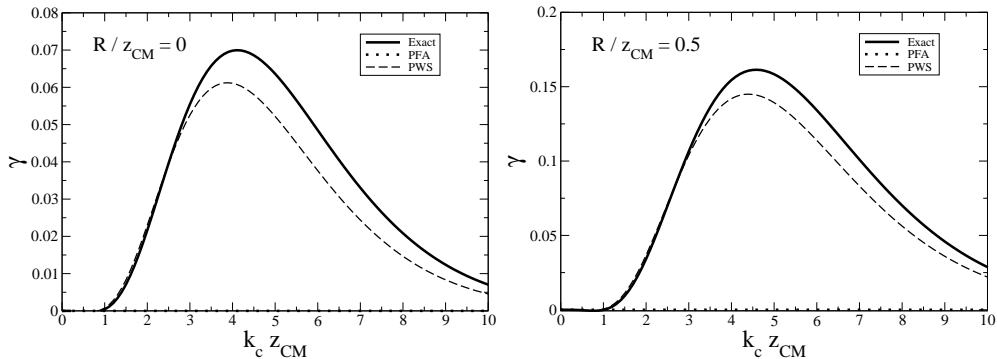

\centering
\includegraphics[width=6.5cm]{fig4_1.eps}
\includegraphics[width=6.5cm]{fig4_2.eps}
\caption{Relative frequency shift due to the lateral Casimir-Polder force on a rubidium BEC 
($m=1.45\times 10^{-25} {\rm kg}$, $\alpha(0)/\epsilon_0=47.3\times 10^{-30} {\rm m}^{3}$) oscillating at a distance $z_{\rm CM}
=2 \mu{\rm m}$ above a perfectly conducting, periodically grooved surface.
The exact result, the proximity approximation, and the paiwise summation approximation are depicted
as a function of $k_c z_{\rm CM}$ for different values of the Thomas-Fermi BEC radius $R$. The unperturbed frequency is $\omega / 2 \pi = 229 {\rm Hz}$, the groove period is $\lambda_c=4 \mu{\rm m}$, the groove width $s=\lambda_c/2$, and the depth $a=250 {\rm nm}$.}
\end{figure}

In Fig. 4 we plot the relative frequency shift $\gamma$ due to the lateral Casimir-Polder force,
and compare the exact result with PFA and PWS. Note that the case $R=0$ corresponds to the single
atom frequency shift $\gamma_0$. As noted before, PFA predicts no frequency shift
since the potential is locally flat above one plateau. Indeed, $\gamma$ is very small for
$k_c z_{\rm CM} < 1$. However, as $k_c z_{\rm CM}$ increases, $\gamma$ develops a peak and then
decays exponentially for $k_c z_{\rm CM} \gg 1$.  PWS underestimates the lateral force,
resulting in a frequency shift smaller than the exact one. 
For the separation chosen in the figure ($z_{\rm CM} = 2 \mu{\rm m}$), the frequency shift
is well above the reported experimental sensitivity $10^{-5}-10^{-4}$ \cite{Cornell2005,Cornell2007}.
However, the results reported in Fig. 4 correspond to the ideal case of a perfectly reflecting plate.
As could be expected, the frequency shift is much lower for real bulk plates. However, the resulting
shifts in those cases are still strong enough to allow for a measurement of the lateral Casimir-Polder
force with a BEC trapped at distances below $z_{\rm CM} = 3 \mu{\rm m}$ \cite{arxiv,future}.
Also, as shown in the figure, the effect of finite size of the BEC is to increase the frequency shift
as the Thomas-Fermi radius $R$ grows.


\section{Conclusion}

The lateral Casimir-Polder force offers the possibility of testing non-trivial
geometrical effects of QED. We have calculated the first order, exact lateral
force on a ground state atom above a corrugated surface and compared it
to the widely used proximity force and pairwise summation approximations.
Our scattering approach allows one to consider real materials and arbitrary
atom-surface separations, as long as the corrugation amplitude remains the smallest
length scale in the problem. On the other hand, the PWS approach needs to be 
``calibrated" for a specific separation distance range since the interaction
is not additive. Even when comparing in a specific distance range (the Casimir-Polder
long distance limit), we have found large deviations from PWS, and even larger deviations
from PFA. Such deviations could be measured using a Bose-Einstein condensate as a 
surface field sensor.


\section{Acknowlegments}

We would like to thank the organizers of QFEXT07.


\end{document}